\documentclass[conference]{IEEEtran}
\IEEEoverridecommandlockouts

\usepackage{cite}
\usepackage{amsmath,amssymb,amsfonts}
\usepackage{algorithmic}
\usepackage{graphicx}
\usepackage{textcomp}
\usepackage{xcolor}
\usepackage{comment}

\def\BibTeX{{\rm B\kern-.05em{\sc i\kern-.025em b}\kern-.08em
    T\kern-.1667em\lower.7ex\hbox{E}\kern-.125emX}}
\begin{document}

\title{Future of Smart Classroom in the Era of Wearable Neurotechnology\\
\thanks{This research was partially supported by NSF awards CISE CRII 2105084 and CMMI 1739503. Any opinions, findings, and conclusions or recommendations expressed in this material are those of the authors and do not necessarily reflect views of our funding agency.}}

\author{\IEEEauthorblockN{Mojtaba Taherisadr, Berken Utku Demirel, Mohammad Abdullah Al Faruque, and Salma Elmalaki}
\IEEEauthorblockA{Department of Electrical Engineering and Computer Science}
\IEEEauthorblockA{University of California, Irvine, California, USA}
\IEEEauthorblockA{{(taherisa, bdemirel, alfaruqu, salma.elmalaki)@uci.edu}}
}

\maketitle

\begin{abstract}

Interdisciplinary research among engineering, computer science, and neuroscience to understand and utilize the human brain signals resulted in advances and widespread applicability of wearable neurotechnology in  adaptive human-in-the-loop smart systems.
Considering these advances, we envision that future education will exploit the advances in wearable neurotechnology and move toward more personalized smart classrooms where instructions and interactions are tailored to students’ individual strengths and needs. 
In this paper, we discuss the future of smart classroom and how advances in neuroscience, machine learning, and embedded systems as key enablers will provide the infrastructure for envisioned smart classrooms and personalized education along with open challenges that are required to be addressed.
\end{abstract}


\section{Introduction}

The disruption of face-to-face exchanges and the urgent digital transformation of our everyday life activities, abide the COVID-19 hit, have spurred us to shift many working paradigms to incorporate the new reality of the online and remote collaborative environment. In particular, many sectors including the education and the healthcare systems had to integrate new technology to assist the continuation of their services even with remote interactions~\cite{taiwo2020smart,mantena2021strengthening}. Nevertheless, this forced digital transformation of many sectors is expected to stay even in the post-COVID-19 era. In particular, in the education sector, the e-learning market worldwide is forecast to surpass 243 billion U.S. dollars by 2022~\cite{elearningStatista}. 

While the education sector has already taken technological leaps over the last decades with the introduction of smart classroom concept~\cite{kwet2020smart,songkram2021developing,dai2020research}, efficient teacher-student interaction has to be envisioned especially with the possibility of remote education. In particular, we envision that the future of education using smart classrooms should move from a \emph{one-size-fits-all} approach to a \emph{personalized} process in which instructions and interactions are tailored to students' individual needs. In this  paper, we argue that advances in bio-sciences, machine learning, and embedded systems can play a significant role in realizing  the vision of personalized and smart education.

\begin{figure*}
\centering
\includegraphics[trim={0cm 9cm 0 0},clip, width=0.99\textwidth]{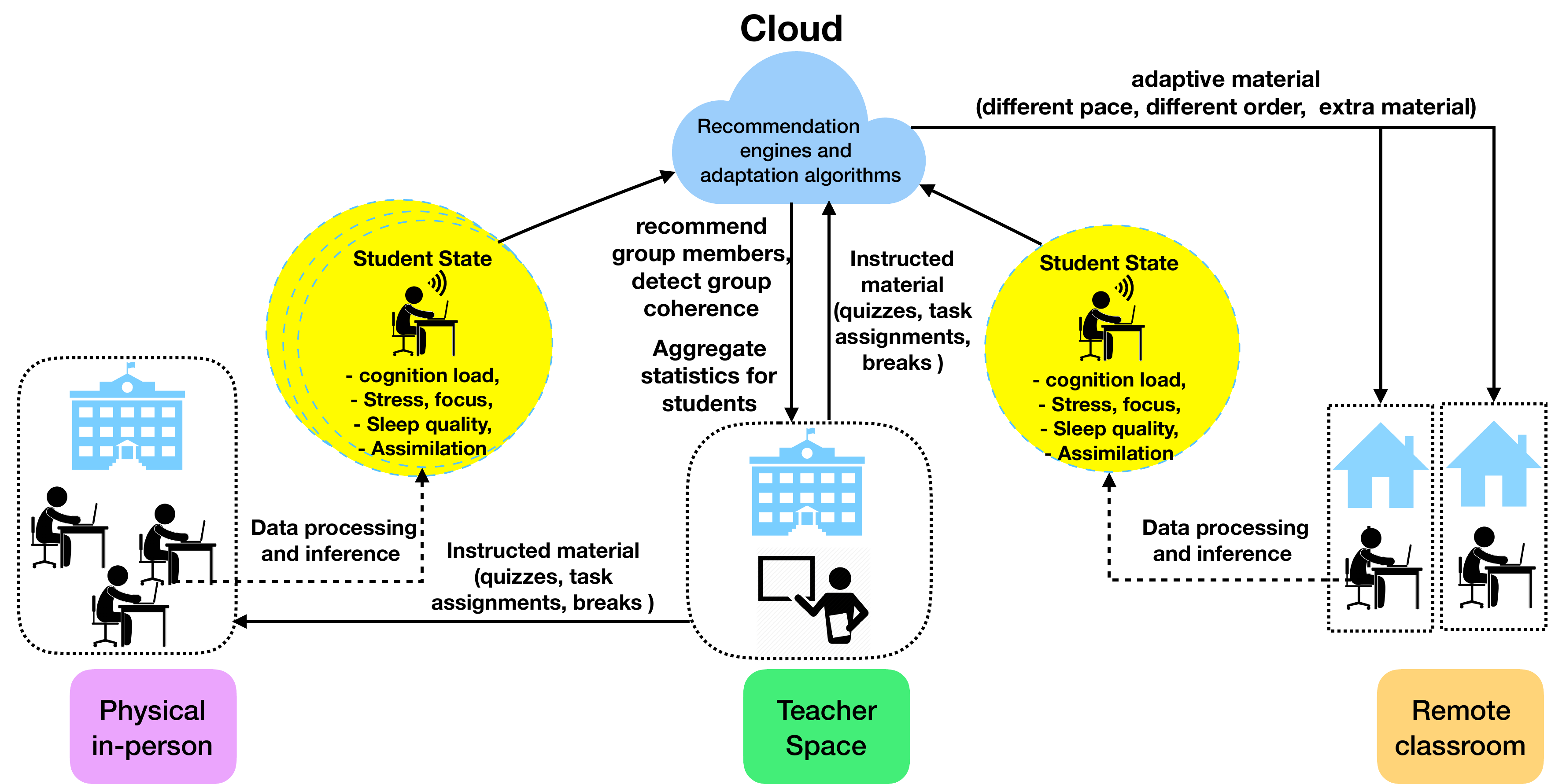}
\caption{Future smart classroom in the era of wearable neurotechnology. Student's mental state inferred from wearable sensors is used to provide feedback and recommendation to the student and the teacher for both physical and remote classes.}
\label{learn}
\end{figure*}

A cornerstone in the envisioned personalized learning is to replace the traditional measures of education (e.g., quizzes, scores in exams, and teacher evaluations) with real-time measurements of the student's state of mind. Such real-time signals can be then used to adapt the instructed materials to maximize the student's performance. The student's stress level, drowsiness level, readiness to learn new knowledge, cognitive load, and learning rate are just examples of such real-time signals that can be used to adapt the instructed materials to enhance the learning performance.
Indeed, there is a tremendous opportunity in attaining this vision. In particular, advances in \textbf{neuroscience} and \textbf{wearable sensors} technology have widen the horizon to reveal the fundamental processes of various human mental states~\cite{chang2021multimodal,gu2021eeg}. These advancements resulted in tremendous research efforts among interdisciplinary fields from engineering researchers, computer scientists, and healthcare (neurologists, clinicians) to integrate the brain state (drowsiness, sleep quality, readiness potential, and learning activity) as an additional sensor modality to design adaptive human-in-the-loop smart systems \cite{elmalaki2021towards}.
Using such advances, we envision the future smart classroom as shown in Figure~\ref{learn}. Future smart classroom will be extended to include physical and remote students. Students' state will be used to adapt the education system through feedback and personalized recommendation to enhance the student's performance.

In what follows, we highlight two scientific and technological advances that are key enablers for the envisioned personalized and smart classrooms along with open challenges that need to be addressed.

\section{Advancement in Neuroscience to understand cognitive processing}
As discussed above, a fundamental challenge to achieve the aforementioned vision is to measure the student's mental state, in real-time. Indeed, recent advances in neuroscience have opened the gate to unveil fundamental processes in the human brain, such as the ability to generate emotions, memories, and actions~\cite{zheng2019multiplexing, ieeebrain}. These research efforts have become possible by the ability to record and stimulate the human brain activity in the clinical setup by neurologists with a very high accuracy~\cite{quiroga2019plugging}. 
Optimal cognitive processing is central to all aspects of human activities, and recent advances in neuroscience have provided critical insights on how the brain accomplishes cognitive processing, spanning single neuron to neural population level resolution~\cite{quiroga2019plugging}.

In particular, recent studies using event-related functional magnetic resonance imaging (fMRI) showed that there exists a high correlation between the magnitudes of focal activation in the right prefrontal cortex and the bilateral parahippocampal cortex during visual learning with the memory processing~\cite{fmri}. In addition, recent findings in literature demonstrating that brain electrophysiological activity observed before stimulus or new concepts presentation profoundly influences subsequent behavior on fear conditioning~\cite{seager2002oscillatory},  motor responses~\cite{mazaheri2009prestimulus}, attention~\cite{makeig2002response}, and memory tasks~\cite{addante2011prestimulus}.

While these studies gave us the fundamentals to measure the ability to memorize visual content, and the specific parts in the brain that are responsible for cognitive processing, the form factor of the utilized machinery prevent the widespread use of this technology in classroom settings. The next section reviews some technological advances that can provide a solution to such a challenge.

\section{Advancement in wearable neurotechnology}

With the development of efficient, intelligent sensing technology, detecting small changes in electric signals becomes plausible. This technology paved the way for developing and producing devices that are capable of monitoring the brain's electrical activity \textbf{non-invasively} with enough resolution and low cost compared to the clinical setups. Although the traditional brain activity monitoring systems suffer from motion artifacts, long preparation time due to conductive gels or pastes, and large equipment (metal electrodes, long wires), the advances in sensor technology allow a fully portable, wireless, long-term, flexible scalp electronic system, incorporating a set of dry electrodes \cite{mahmood}. Moreover, the importance of wearable brain activity monitoring devices is that they are feasible for envisioning a real-time human-in-the-loop education system thanks to their portability, comfortability, and wireless data transfer system. 

A study by Dikker et al. \cite{Classroom} proves that using a portable electroencephalogram (EEG) device for recording brain activity from a class of 12 high school students over a semester during regular activities, analyzing the group-based neural coherence is possible where the brain activity is synchronized across students in both student class engagement and social dynamics.

Another study by Babini et al.\cite{vrtech} comparatively measured the learning of the students in a virtual reality (VR) environment for using a wearable electroencephalogram (EEG)~\cite{EMOTIV}. In particular, they compared the brain engagement level of a group of students while learning through a VR environment compared to the regular two-dimensional environment. Their study showed that the higher the brain engagement, the higher the attention level of the students which led to  better learning performance.

Moreover, the developments in wearable neurotechnology devices 
can enable complex applications to be integrated to understand and predict  cognitive processing. For example, sleep has a pivotal role in cognitive functions, and their relationship has been a topic of interest for over a century. Extensive research on sleep studies has shown that better sleep is associated with a myriad of superior cognitive functions in healthy adults \cite{memory1,memory2}, including better learning and memory \cite{learning1}. Although the exact mechanisms behind the relationship between sleep, memory, and learning are still research topics, the general agreement is those specific synaptic connections that were active during awake periods are strengthened during sleep, allowing for the consolidation of memory, and inactive synaptic connections are weakened \cite{learning2}. Consequently, sleep provides an essential function for memory consolidation; in other words, it will enable us to remember the topics that have been studied before\cite{learning2}. It is now well established from a variety of studies that sleep monitoring can be used to associate with the students' performances. Moreover, sleep monitoring should be long-term rather than a single night or particular event recordings. Therefore, the recent developments on low-power wearable devices have crucial importance in monitoring students' sleep or emotional situations in the long term to create the future smart classroom environments~\cite{demirel2021singlechannel}.

\section{Personalized feedback for the future smart classroom}

When these advancements in neuroscience and sensor technology combined with the current evolution in machine learning and signal processing, classification, and analysis of the massive amount of data streaming from wearable devices have become practicable. The resourcefulness of machine learning models provides the ability to analyze the complex recurring patterns of neural activity and detect hidden patterns. However, there are several challenges that need to be addressed to use these advancements to provide personalized feedback to the students and adapt the teaching environment for  future smart classrooms.

\begin{itemize}
    \item \textbf{Challenge 1: Decision making in the face of system variability:}
    Even if we are able to decode the brain activity in real-time to estimate the cognitive processing, addressing the students' individual needs to achieve personalized feedback has a lot of challenges. In particular, there are intrinsic variations that humans exhibit that can complicate the idea of a personalized smart classroom. As discussed in one of our recent works, the same human's preferences and responses may change over time. Even for a small period of time, the same human may produce different responses based on unmodeled external effects (intra-human variability). Similarly, different humans may have different responses under similar conditions (inter-human variability)~\cite{elmalaki2018sentio}. For example, adapting the pace of the teaching material or the time of a quiz based on the cognitive load and the sleep quality of the student may not be the same for a different student and even for the same student across time. Moreover, when multiple humans interact in the same environment, their preferences differ based on the number of people they interact with and the type of this interaction. Moreover, when the education system adapts to one student in a multi-student classroom (e.g., change the pace of the presented material), it has a direct effect on the other students in the same classroom. In addition to that, multiple students may require different feedback adaptation actions based on their different preferences (multi-human variability) which can raise a question of fairness of adaptation in future smart classrooms~\cite{elmalaki2021fair, elmalaki2018internet}. Hence, these human variations have to be considered as an integral part of the design process of the adaptation algorithms for the future of the smart classroom.
    
    \item \textbf{Challenge 2: Restructuring the pedagogical materials:} 
    Prior to applying the vision of personalized feedback in the future smart classroom, the pedagogical materials have to be reconstructed to be adaptable. For example, teaching materials have to be redesigned into small modules that are self-contained to enable quizzes and breaks at any time. Moreover, the same teaching materials themselves have to be designed into different modalities, such as project-based learning with a small group of peers, independent work to complete complex tasks, using immersive technology (VR) to increase the engagement of some students, and even recordings of the material to be watched at a different pace. Quizzes have to redesigned be automated based on the time they are taken and the student's strengths. Projects and group assignments have to take into account the group coherency and the average learning performance of the group members. Even teachers have to be trained to keep an up-to-date record from real-time data that provides a deep understanding of each student's individual strengths, needs, motivations, progress and goals~\cite{morin2014everything}. 
        
    \item \textbf{Challenge 3: Internet-of-Things (IoT) design and robustness to errors:} 
    Infusing technology into pedagogy to attain the vision of the future smart classroom requires careful design of the enabling technology to ensure high accuracy of real-time data and robustness to errors. Albeit the wide-ranging adaptation of the Internet-of-Things (IoT) technology in different application domains, low power and low energy design of wearable embedded systems are some of the most critical challenges to realize this new paradigm of future smart classrooms and to ensure its feasibility. Moreover, the need to process EEG data at the edge level with high accuracy (popularly termed as machine learning at the edge) due to privacy, security, and communication latency require new hardware-software design techniques to achieve low cost for accurate real-time local processing~\cite{demirel2021singlechannel}. Current research in the area of battery-aware task scheduling for wearables~\cite{elmalaki2014case}, post CMOS technology, and Neuromorphic computing systems~\cite{computing2015neuromorphic,roy2019towards} will play a significant role in developing such low power, low energy, and high accuracy efficient IoT design for future smart classrooms. 
    
\end{itemize}

\section{Conclusion}
This is an opportune time for smart classrooms. The increased demands of new affordable yet accurate solutions for human-in-the-loop systems lead to rich research areas, especially in human sensing. With the advancement in neuroscience and wearable neurotechnology, new possibilities for improving the quality of the education system by merging both fields have emerged. However, new challenges are yet to be solved in all development phases of future smart classrooms, from more neuroscience insights into the brain circuitry to design phases of accurate sensing to efficient adaptation algorithms. This article highlights the most prominent opportunities for future smart classrooms and recent approaches to tackle them. Nevertheless, this is not an exhaustive list leaving much to be discovered and done.

\bibliographystyle{IEEEtran}
\bibliography{refs}

\end{document}